**Efficient spin current source using a half-Heusler alloy topological semimetal with Back-End-of-Line compatibility**

Takanori Shirokura[1], Tuo Fan[1], Nguyen Huynh Duy Khang[1,a], Tsuyoshi Kondo[2], and Pham Nam Hai[1,b]*

[1] *Department of Electrical and Electronic Engineering, Tokyo Institute of Technology, Meguro, Tokyo 152-8550, Japan*

[2] *Device Technology R&D Center, Institute of Memory Technology R&D, Kioxia Corporation, Yokohama, Kanagawa 235-0032, Japan*

a) On leave from Department of Physics, Ho Chi Minh City University of Education

b) Also at Center for Spintronics Research Network (CSRN), The University of Tokyo

*Corresponding author: pham.n.ab@m.titech.ac.jp

**Abstract**

Topological materials, such as topological insulators (TIs), have great potential for ultralow power spintronic devices, thanks to their giant spin Hall effect. However, the giant spin Hall angle ($\theta_{SH} > 1$) is limited to a few chalcogenide TIs with toxic elements and low melting points, making them challenging for device integration during the silicon Back-End-of-Line (BEOL) process. Here, we show that by using a half-Heusler alloy topological semi-metal (HHA-TSM), YPtBi, it is possible to achieve both a giant $\theta_{SH}$ up to 1.6 and a high thermal budget up to 600°C. We demonstrate magnetization switching of a CoPt thin film using the giant spin Hall effect of YPtBi by current densities lower than those of heavy metals by one order of magnitude. Since HHA-TSM includes a group of three-element topological materials with great flexibility, our work opens the door to the third-generation spin Hall materials with both high $\theta_{SH}$ and high compatibility with the BEOL process that would be easily adopted by the industry.

**Introduction**

Pure spin current induced by the spin Hall effect in materials with strong spin-orbit coupling is a promising magnetization manipulation method for various next-generation spintronic devices, such as spin-orbit torque magnetoresistive random-access memories (SOT-MRAM)[1,2], race track memories[3], and spin Hall oscillators[4,5]. There are two families of materials that are currently studied for efficient spin current sources: heavy metals (HMs) and topological insulators (TIs). HMs, such as Ta,[6] W,[7] Pt,[8] and their alloys with other elements, have the advantages of high melting point and non-toxicity. Furthermore, some of them have been already adopted as buffering or lining materials in silicon Back-End-of-Line (BEOL) process. Thus, these HMs, considered as the first-generation spin Hall materials, have been heavily studied by the industry as candidates for spin current sources in spintronic devices. However, the spin Hall performance of HMs is insufficient because their spin Hall angle $\theta_{SH}$ is usually smaller than 1. On the hand, TIs with topological surface states (TSS), such as $Bi_2Se_3$, $(Bi,Sb)Te_3$, and BiSb, have demonstrated very high $\theta_{SH}$ larger than 1 at room temperature in epitaxial TI thin films prepared by molecular beam epitaxy.[9,10,11] Moreover, the high $\theta_{SH}$ is maintained even in non-epitaxial TIs prepared by the industry-friendly sputtering technique.[12,13,14,15] Thus, TIs are very promising for magnetization manipulation with ultralow power consumption, and considered as the second-generation of spin Hall materials. However, the observed room-temperature high $\theta_{SH}$ is so far limited to only a few chalcogenide TI materials with toxic elements of either Se, Sb, or Te. Furthermore, those TIs have low melting points, making them challenging for device integration during the silicon BEOL process, which involves high temperature up to 400°C. As a result, despite having very high $\theta_{SH}$, the TI-based second-generation spin Hall materials have not yet been adopted by the industry as candidates for spin current sources. Although there are recent attempts to investigate the spin Hall effect in other topological materials beyond TIs, such as Weyl semimetals,[16,17] none of them has shown high $\theta_{SH}$ (>1) and achieved BEOL compatibility that are essential for industrial spintronic applications.

In this work, we demonstrate a spin Hall material that inherits the advantage of HMs and TIs, using a half-Heusler alloy topological semimetal (HHA-TSMs), YPtBi. We demonstrate that YPtBi can have a high $\theta_{SH}$ up to 1.6, rivaling that of TIs, and a high thermal budget up to 600°C, comparable to that of HMs. We show that the high $\theta_{SH}$ can be explained by the spin Hall effect of TSS rather the bulk states. We then demonstrate magnetization switching of a CoPt thin film using the giant spin Hall effect of YPtBi by current densities lower than those of heavy metals by one order of magnitude. Since HHA-TSM includes a group of three-element topological materials with great flexibility, our work opens the door to the third-generation spin Hall materials with both high $\theta_{SH}$ and high compatibility with the BEOL process that would be easily adopted by the industry.

**Comparison between TI and HHA-TMS**

Figure 1 compares the crystal structure and the band structure of TI and HHA-TSM. Most TIs crystallize in the trigonal or rhombohedral lattices including two dimensional atomic sheets interacting via Van der Waals bonding, as schematically shown in the left panel of Fig. 1(a). While this crystal structure makes it easy to grow high-quality TIs on many substrates, it results in low mechanical strength and low melting points that in turn increase the crystal grain and surface roughness, which are not favored for device integration. The schematic band structure of TIs is shown in the right panel of Fig. 1(a), which consists of a bulk band gap and one or multi TSS with Dirac-like dispersion and spin-momentum locking. The large Berry phase curvature originating from the monopole at the Dirac cones of TSS is the key for obtaining the giant spin Hall angle in TIs. The left panel in Fig. 1(b) shows the schematic crystal structure of HHA-TSMs, which are constructed from three kinds of element XYZ, where X and Y are transition or rare-earth metals, and Z is the main-group element, and thus possesses high controllability of lattice constant and band structure via material combination.[18,19,20,21,22] Unlike TIs, the crystal structure of HHA-TSMs are cubic, matching those of many ferromagnetic materials and the MgO insulating material typically used in realistic spintronic devices. Furthermore, HHA-TSMs have high melting points, which make them compatible with the BEOL process. In numerous combinations of X, Y, and Z, an inversion of the s-orbital band with $\Gamma_8$ symmetry and the p-orbital band with $\Gamma_6$ symmetry occurs around the $\Gamma$ point, and then TSS is generated through the band topology change,[23,24] as schematically shown in the right panel of Fig. 1(b). An ideal topological HHA would have zero band gap and no Fermi surface. In reality, due to crystal defects, there can exist a significant number of carriers and thus non-zero Fermi surface, making topological HHA semi-metal rather than zero-gap insulator. Numerous works have successfully observed such a Fermi surface and TSS in several HHA such as LuPtBi, YPtBi, and LuPtSb using angle-resolved photoelectron spectroscopy,[25,26,27,28] confirming that they are HHA-TSMs. Nevertheless, there is no report on the spin Hall performance of these compounds.

**Deposition and characterization of single YPtBi thin films**

Here, we chose to investigate the spin Hall effect in YPtBi as a proof of concept for several reasons. First, the band inversion between $\Gamma_8$ and $\Gamma_6$ of YPtBi are among the largest as predicted from first principle calculation.[18,19] Furthermore, this compound does not contain toxic elements, such as Pb,Th, or Sb. Finally, Y is stable in air and is easy to handle than other rare-earth elements, such as Lu, Ce, and La. We grew YPtBi thin films on c-sapphire substrate by co-sputtering multi targets (see Methods). In this section, to demonstrate the high thermal stability of YPtBi, we prepared two different series of samples with different substrate temperature and Ar gas pressure. The sample structure is $MgAl_2O_4$ (2.0)/YPtBi (~50)/c-Sapphire, where the layer thicknesses are in nm. The cap $MgAl_2O_4$ (2.0) layer was deposited at room temperature. Figure 2(a) shows the X-ray diffraction (XRD) $\theta$-$2\theta$ spectra for YPtBi films deposited at different substrate temperature $T_S$ ranging from 300ºC to 800ºC and the Ar pressure of 2.0 Pa. Clear YPtBi(111) peaks were observed at $T_S$ = 300ºC ~ 600ºC, indicating that YPtBi is stable up to 600ºC. The lattice constant of 6.62 Å evaluated from the peak position of YPtBi(111) is consistent with the bulk value of 6.64 Å.[29] Therefore, a strain effect on the band structure is negligibly small in our YPtBi films. Figure 2(b) shows the XRD $\theta$-$2\theta$ spectra for YPtBi films deposited at different Ar pressure ranging from 0.3 to 2.0 Pa at $T_S$ = 600ºC. Peaks of YPtBi(111) were observed under the whole Ar pressure range. The inset in Fig. 2(b) shows the peak intensity $I_N$ of YPtBi(111) normalized by that at the Ar pressure of 2.0 Pa. $I_N$ increases with increasing Ar pressure up to 1 Pa, above which $I_N$ saturates because higher Ar pressure reduces the recoil energy of Ar ions which may implant into the YPtBi thin films and reduce their crystal quality. We then used X-ray fluorescence spectroscopy (XRF) to characterize the elemental composition of the YPtBi thin film. A representative XRF spectrum of the YPtBi thin film deposited at $T_S = 600°C$ is shown in Fig. 2(c). Fitting to the intensity of the characteristic X-ray energy of each element allows us to determine that the atomic composition Y:Pt:Bi of this sample is close to 1:1:1. Figure 2(d) shows the relative atomic composition of Bi as a function of $T_S$. Since Bi is the most volatile among the three elements, this data allows us to determine whether YPtBi is stable or not at a particular substrate temperature. The result in Fig. 2(d) reveals that YPtBi is stable up to 600°C, above which Bi composition significantly decreases due to desorption from YPtBi, consistent with the XRD spectra in Fig. 2(a). There results demonstrate that YPtBi has a large thermal budget up to 600°C, higher than the required 400°C of the BEOL process. This is a significant advantage of HHA-TSMs compared with TIs.

To investigate the electric properties of YPtBi, we prepared a thinner $MgAl_2O_4$ (2.0)/ YPtBi (11.8)/c-sapphire stand-alone sample. We fabricated a 4-terminal Hall bar structure with size of 60×100 $\mu m^2$ by optical lithography and ion-milling for electrical measurements. We obtained the charge conductivity $\sigma_{YPtBi}$ of $1.5×10^5$ $\Omega^{-1}m^{-1}$, which is similar to that of bulk YPtBi.[30] From the Hall measurement, we confirmed a large carrier density of $7.1 \times 10^{22}$ cm$^{-3}$. Figure 3(a) shows the temperature dependence of the resistivity $\rho_{\text{YPtBi}}$ of this YPtBi film. We observed that $\rho_{\text{YPtBi}}$ increases with lowering temperature, consistent with the semi-metallic behavior observed in bulk YPtBi.[30] However, the change of $\rho_{\text{YPtBi}}$ at low temperature is only 4%, which is one order of magnitude smaller than that (~50%) observed in bulk YPtBi. This can be explained by the increasing contribution of metallic surface conduction at small thickness. We then employ the planar Hall effect measurement to detect the existence of the spin-momentum locking of TSS in YPtBi. Figure 3(b) shows the DC planar Hall resistance measured at 4 K with an in-plane rotating magnetic field $H_{\text{ext}}$ of 8.4 kOe. Here, dots show experimental data and the solid curve is a fitting result by[31],

$$R_{xy}^{\text{DC}} = R_{\text{PHE}}^{\text{TSS}} \sin 2\phi + R_{\text{OHE}}^{x} \cos \phi + R_{\text{OHE}}^{y} \sin \phi, \quad (1)$$

where $R_{\text{PHE}}^{\text{TSS}}$ is a planar Hall resistance originated from spin-momentum locking of TSS, $R_{\text{OHE}}^{x}$ and $R_{\text{OHE}}^{y}$ are an ordinary Hall resistance caused by the misalignment of $H_{\text{ext}}$ from the *xy* plane, $\phi$ is the azimuth angle for $H_{\text{ext}}$, as shown in the inset. We observed a clear $\sin 2\phi$ planar Hall effect despite YPtBi is a non-magnetic material. Such a $\sin 2\phi$ planar Hall effect has been reported in many non-magnetic TIs with TSS crossing the Fermi level.[31,32,33,34] Because $H_{\text{ext}}$ breaks the time-reversal-symmetry of TSS, back scattering increases due to selective destruction of spin-momentum-locking along the $H_{ext}$ direction, resulting in the $\sin 2\phi$ planar Hall effect.[31] To further confirm this scenario, we shows in Fig. 3(c) the amplitude $R_{PHE}^{TSS}$ measured at various $H_{ext}$, where dots are experimental data and the solid line is a fitting result by a quadratic function. The observed $R_{PHE}^{TSS} \sim H_{ext}^2$ is consistent with previous reports on the planar Hall effect of TIs,[31-34] which provide an evidence that our YPtBi films have TSS crossing the Fermi level. Thus, a large spin Hall effect from the Berry phase curvature of TSS can be expected.

**Charge-to-spin conversion efficiency of YPtBi**

We now investigate the spin Hall angle $\theta_{SH}$ of YPtBi in three samples with different $\sigma_{YPtBi}$ of 0.38, 1.2, and 1.5×10$^5$ $\Omega^{-1}$m$^{-1}$ (referred below as sample A, B, C). We prepared multilayers of MgAl$_2$O$_4$ (2.0)/Pt (0.5)/Co ($t_{Co}$)/Pt (0.5)/YPtBi ($t_{YPtBi}$)/c-sapphire, as shown in Fig. 4(a). The YPtBi layers were deposited by co-sputtering YPt and Bi targets. The Pt/Co/Pt ferromagnetic multilayers have perpendicular magnetic anisotropy (PMA), and referred below as CoPt for short. Here, the parasitic spin Hall effect from the Pt layers is negligible because the Pt thickness of 0.5 nm is few times to one magnitude thinner than the typical spin relaxation length of Pt, and the Pt/Co/Pt stack is symmetric.[14,35,36,37] The thickness of the Co layer $t_{Co}$ is 0.5 nm for sample A and B, and 0.8 nm for sample C. The YPtBi layer thickness $t_{YPtBi}$ for sample A, B, and C is 9.3, 11.5 and 11.8, respectively, as measured by X-ray reflectivity for stand-alone YPtBi thin films deposited at the same conditions. After the deposition, these film stacks were patterned into 25×50 μm² Hall bar devices for transport measurements. Figures 4(b) and 4(c) show the DC anomalous Hall resistance for sample A measured at room temperature with $H_{ext}$ applied along the z-direction and x-direction. Strong perpendicular magnetic anisotropy (PMA) with the effective perpendicular magnetic anisotropy field $H_k^{eff}$ of 6.1 kOe indicates that YPtBi can provide a very flat interface, which is an advantageous feature of YBiPt comparing with TIs for use in SOT devices. By tuning the growth condition, we can obtain atomically flat surface of YPtBi with the surface roughness of 2.4 Å, i.e. less than one atomic layer (see Supplementary Information), which is surprising given that the YPtBi thickness is ~ 10 nm. This flat surface helps induce a large PMA of the Pt/Co/Pt stack on top. For comparison, the surface roughness of 10 nm-thick BiSb TI deposited on c-plane sapphire is 6 Å, and it is very difficult to obtain PMA for Pt/Co/Pt on top of BiSb due the large surface roughness. Quantitative evaluation of $\theta_{SH}$ in sample A was carried out by using the high-field second harmonic technique at room temperature with alternating currents at 259.68 Hz. In the case of PMA, the second harmonic Hall resistance $R_{xy}^{2\omega}$ measured at $H_{ext}$ higher than $H_k^{eff}$ applied along the x-direction is given

by,[38,39]

$$R_{xy}^{2\omega}=\frac{R_{AHE}}{2}\frac{H_{DL}}{|H_{ext}|-H_k^{eff}}+R_{PHE}\frac{H_{FL+OF}}{|H_{ext}|}+\alpha_{ONE}|H_{ext}|+R_{ANE+SSE}, \quad (2)$$

where $H_{DL}$ is the antidamping-like field, $H_{FL+OF}$ is the sum of the field-like and Oersted field, $R_{AHE}$ is the anomalous Hall resistance, $R_{PHE}$ is the planar Hall resistance, $\alpha_{ONE}$ is a coefficient reflecting contribution from the ordinary Nernst effect, and $R_{ANE+SSE}$ is a constant reflecting contribution from the anomalous Nernst effect and the spin Seebeck effect. Here, fitting parameters are $H_{DL}$, $H_k^{eff}$, $H_{FL+OF}$, $\alpha_{ONE}$, and $R_{ANE+SSE}$ that can be determined uniquely, according to linear algebra. We note that this high-field second harmonic technique can distinguish the SOT effect from the thermal effects thanks to their different $H_{ext}$-dependence, while it is difficult to do so for the low-field second harmonic technique where the SOT and thermal effects have the same linear $H_{ext}$-dependence. Representative high-field second harmonics data and the corresponding fitting for sample A at bias currents of 1.0 to 3.4 mA are shown in Fig. 4(d), where the dots and solid curves are the experimental data and fitting using equation (2), respectively. Figure 4(e) shows the relationship between the extracted values of $H_{DL}$ and the current density in the YPtBi layer $J_{YPtBi}$ for sample A. Then, the effective spin Hall angle $\theta_{SH}^{eff}$ was calculated from the slope of $H_{DL}/J_{YPtBi}$ by,

$$\theta_{SH}^{eff}=\frac{2eM_s t_{CoPt}}{\hbar}\frac{H_{DL}}{J_{YPtBi}}, \quad (3)$$

where $e$ is the electron charge, $\hbar$ is the Dirac constant, $M_S = 633$ emu/cc is the saturation magnetization of the CoPt layer measured by superconducting quantum interference device (SQUID), and $t_{CoPt} = 1.5$ nm is the thickness of the CoPt layer. Thanks to the contribution of TSS, large $\theta_{SH}^{eff}$ of 1.3 was observed in sample A. Furthermore, our fittings to high-field second harmonic data indicate that the thermal contribution is negligible in sample A, thus the low-field second harmonic technique[38] can also be used to double check the value of $\theta_{SH}^{eff}$, which indeed confirms $\theta_{SH}^{eff} = 1.3$ (see Supplementary Information). This large $\theta_{SH}^{eff}$ is comparable with those reported in TIs such as $Bi_2Se_3$ and $(BiSb)_2Te_3$,[9,10] and larger than that of topological Weyl semimetals.[16,17] (For a sample E consisting of an YPtBi layer with the same conductivity but deposited by co-sputtering Y, Pt, and Bi target, $\theta_{SH}^{eff}$ is even larger, reaching 1.6 as shown in the next section)

We then investigate the spin Hall effect in sample B and C with higher $\sigma_{YPtBi}$ of 1.2 and $1.5 \times 10^5$ $\Omega^{-1}m^{-1}$, respectively. In sample B, $H_k^{eff}$ is as strong as 7.4 kOe for that we cannot obtain enough high-field second harmonic data for fitting. To overcome this problem, we employed the angle-resolved second harmonic technique to separate the contribution from SOT and thermal effects,[40] from which we obtained $\theta_{SH}^{eff} = 0.9 \pm 0.2$ in sample B (see Supplementary Information). For the sample C with higher $\sigma_{YPtBi}$ of $1.5 \times 10^5$ $\Omega^{-1}m^{-1}$ than that of sample B, we reduced $H_k^{eff}$ to 6.1 kOe by increasing the Co thickness to 0.8 nm and evaluated $\theta_{\text{SH}}^{\text{eff}}$ by the high-field second harmonic technique, which yields $\theta_{\text{SH}}^{\text{eff}} = 0.64$ (see Supplementary Information). Figure 4(f) summarizes the $\theta_{\text{SH}}^{\text{eff}} - \sigma_{\text{YPtBi}}$ relation for each sample. Here, $\theta_{\text{SH}}^{\text{eff}}$ for sample E which is discussed in the next section is also plotted. We model the spin Hall angle by considering the contribution from a TSS spin Hall conductivity $\sigma_{\text{SH}}^{\text{TSS}}$ and a bulk spin Hall conductivity $\sigma_{\text{SH}}^{\text{B}}$,

$$\theta_{\text{SH}}^{\text{eff}} = \frac{\sigma_{\text{SH}}^{\text{TSS}} + \sigma_{\text{SH}}^{\text{B}}}{\sigma_{\text{YPtBi}}}. \qquad (4)$$

Thus, $\theta_{\text{SH}}^{\text{eff}} \sim \sigma_{\text{SH}}^{\text{TSS}}/\sigma_{\text{YPtBi}}$ if TSS dominates the spin Hall effect, and $\theta_{\text{SH}}^{\text{eff}} \sim \sigma_{\text{SH}}^{\text{B}}/\sigma_{\text{YPtBi}}$ if otherwise. Regardless of which dominates, because the spin Hall conductivity is constant in the moderately dirty regime with the dominating Berry phase mechanism,[41,42] $\theta_{\text{SH}}^{\text{eff}}$ is inversely proportional to $\sigma_{\text{YPtBi}}$. The red solid curve in Figure 4(f) is a fitting result of the $\theta_{\text{SH}}^{\text{eff}} - \sigma_{\text{YPtBi}}$ relation by equation (4), which reasonably agrees with the experimental results and indicates that the spin Hall effect in YPtBi is indeed governed by the Berry phase.

In order to see which of $\sigma_{\text{SH}}^{\text{TSS}}$ or $\sigma_{\text{SH}}^{\text{B}}$ dominates the spin Hall effect, we studied the YPtBi thickness dependence of $\theta_{\text{SH}}^{\text{eff}}$ in another series of samples with structure of Ta (1.0) / MgAl$_2$O$_4$ (2.0)/Pt (0.8)/Co (0.5)/Pt (0.8)/YPtBi ($t_{\text{YPtBi}}$)/c-sapphire, whose $t_{\text{YPtBi}}$ = 4, 6, 8, 10, 20 nm. The YPtBi layers in this series were deposited by co-sputtering Y, Pt, and Bi targets. A representative sample D with $t_{\text{YPtBi}}$ = 10 nm and high $\sigma_{\text{YPtBi}}$ = 2.0×10$^5$ $\Omega^{-1}\text{m}^{-1}$ shows $\theta_{\text{SH}}^{\text{eff}} = 0.12$ (see Supplementary Information). The inset in Fig. 4(f) shows the thickness dependence of $\theta_{\text{SH}}^{\text{eff}}$ in this series, where the dots and solid line indicate the experimental data and fitting result by equation (4). At $t_{\text{YPtBi}} \geq 8$ nm, equation (4) explains the experimental results very well. At $t_{\text{YPtBi}} < 8$ nm, $\theta_{\text{SH}}^{\text{eff}}$ rapidly decreases to below that expected from the equation (4). Notably, $\theta_{\text{SH}}^{\text{eff}}$ reaches 0 at $t_{\text{YPtBi}} = 4$ nm. This behavior cannot be explained by the bulk spin Hall effect, whose thickness dependence should follow $\theta_{\text{SH}}^{\text{eff}} = \theta_{\text{SH}}^{\text{B}}\left[1 - \text{sech}\left(\frac{t_{\text{YPtBi}}}{\lambda_{\text{S}}}\right)\right]$, which means $\theta_{\text{SH}}^{\text{eff}} \sim \theta_{\text{SH}}^{\text{B}}$ at $t_{\text{YPtBi}} \gg \lambda_{\text{S}}$ and $\theta_{\text{SH}}^{\text{eff}} \sim \frac{1}{2}\theta_{\text{SH}}^{\text{B}}\left(\frac{t_{\text{YPtBi}}}{\lambda_{\text{S}}}\right)^2$ at $t_{\text{YPtBi}} \ll \lambda_{\text{S}}$, where $\lambda_{\text{S}}$ is the spin diffusion length. Thus, the bulk $\theta_{\text{SH}}^{\text{eff}}$ should approach zero only at $t_{\text{YPtBi}} = 0$ nm. Meanwhile, if $\theta_{\text{SH}}^{\text{eff}}$ is governed TSS, the observed rapid decrease and disappearance of $\theta_{\text{SH}}^{\text{eff}}$ at $t_{\text{YPtBi}} = 4$ nm can be explained by destruction of spin Hall current due to interference of TSS on the top and bottom surface of YPtBi, a phenomenon well known in TIs.[43] Thus, our results indicate that the spin Hall effect in YPtBi is dominated by the intrinsic mechanism (Berry phase) from TSS.

**Magnetization switching by ultralow DC and pulse currents**

To demonstrate the magnetization switching by the SOT effect generated by YPtBi, we prepared a stack of Ta (1.0)/MgAl$_2$O$_4$ (2.0)/Pt (0.8)/Co (0.5)/Pt (0.8)/YPtBi (11.3)/c-sapphire (sample E). We fabricated a Hall bar device with size of 10×60 μm$^2$ of this stack for current-induced magnetization switching. In sample E, the CoPt layer has $H_k^{\text{eff}}$ of 3.4 kOe. $\sigma_{\text{YPtBi}}$ was tuned to $0.39 \times 10^5$ $\Omega^{-1}\text{m}^{-1}$ to maximize $\theta_{\text{SH}}^{\text{eff}}$, which is estimated to be 1.6 from high-field second harmonic data (see Supplementary Information). Figure 5(a) shows the SOT magnetization switching loops at room temperature by DC currents under an in-plane bias magnetic field $H_{\text{ext}}$ (-1.4 ~ 1.4 kOe) applied along the x-direction. The switching polarity was reversed when $H_{\text{ext}}$ with opposite direction was applied, indicating the switching was governed by SOT. Furthermore, the observed switching direction shows that the spin Hall angle of YPtBi has the same polarity with that of Pt,[44] and is similar to that of other Bi-based TIs.[9-11]

Figure 5(b) shows the threshold current density $J_{\text{th}}^{\text{YPtBi}}$ as a function of $H_{\text{ext}}$. Low $J_{\text{th}}^{\text{YPtBi}}$ on the order of $10^5$ Acm$^{-2}$ was achieved for entire $H_{\text{ext}}$. Figure 5(c) shows the full SOT switching loops by pulse currents with various pulse width $\tau$ from 50 μs to 10 ms under $H_{\text{ext}}$ of 0.5 kOe applied along the x-direction. Figure 5(d) shows $J_{\text{th}}^{\text{YPtBi}}$ as a function of $\tau$. For a reasonable benchmarking, we note that the switching current density in 10×60 μm$^2$ Hall bar of Pt (3)/[Co(0.4)/Pt(0.4)]$_n$ stacks is $3.2 \times 10^7$, $4.6 \times 10^7$, and $5.4 \times 10^7$ Acm$^{-2}$ at the pulse width of 100 ms for n = 1, 2, 4, respectively.[45] Thus, the switching current density in YPtBi is one order of magnitude lower than that of Pt, demonstrating that YPtBi is as efficient as TIs.

We then fit $J_{\text{th}}^{\text{YPtBi}}$ by the thermal fluctuation model, [46,47]

$$J_{\text{th}}^{\text{YPtBi}} = J_{\text{th0}}^{\text{YPtBi}} \left[1 - \frac{1}{\Delta} \ln\left(\frac{\tau}{\tau_0}\right)\right], \quad (5)$$

where, $J_{\text{th0}}^{\text{YPtBi}}$ is the threshold current density for YPtBi layer at 0 K, $\Delta$ is the thermal stability factor, and $1/\tau_0$ (10 GHz) is the attempt frequency associated with the precession frequency of a magnetization. From the fitting, we obtained $\Delta$ = 39 and $J_{\text{th0}}^{\text{YPtBi}}$ = $1.8 \times 10^6$ Acm$^{-2}$. Here, $\Delta$ is similar to that of CoFeB/MgO in perpendicular magnetic tunnel junctions with diameter of 20 ~ 30 nm. Finally, to demonstrate robust SOT switching using YPtBi, we applied a sequence of 105 pulses with $J_{\text{YPtBi}}$ of 1.7 MA/cm$^2$ and $\tau$ of 50 μs, as shown in Fig. 6(a). Figure 6(b) shows the Hall resistance data measured for a total of 210 pulses under the bias field of $H_{\text{ext}} = \pm 0.5$ kOe. We observed very robust SOT switching by YPtBi, indicating that the spin Hall characteristics of YPtBi were unchanged during the pulse sequences.

**Discussion**

We have demonstrated a proof of concept that by using a HHA-TSM, we can combine the advantage of the high spin Hall performance of TIs ($\theta_{\text{SH}}^{\text{eff}} > 1$) and the high thermal stability (> 400°C) of HMs, which make it much easier for industrial adoption than the case of TIs. Our results open the door to the third-generation spin Hall materials with both high $\theta_{SH}$ and high compatibility with the BEOL process. Since HHA-TSMs include a group of three-element topological materials with great flexibility for material choice, we call for further investigation of the spin Hall performance of this family. Indeed, first principle calculations have suggested at least ten materials in this group.[18,19] Although the YPtBi thin films in our work were deposited using multi target sputtering, bulk single crystal of some bismuthides, such as HoPdBi, LuPdBi, LuPtBi, YPtBi, GdPdBi, and DyPdBi have been synthesized by the Bi-flux technique,[30] indicating that it is possible to make large single crystal targets of HHA-TSM for mass production. Contrasting to the case of TIs, we found that there is no obvious physical nor technical difficulty for HHA-TSMs to be used in realistic spintronic applications.

**Method**

**Material growth**

We deposited YPtBi on c-sapphire substrates by co-sputtering multi targets (stoichiometric YPt alloy and Bi for sample A-C, and Y, Pt, Bi for sample D and E). The sputtering condition (Ar pressure, substrate temperature, power, etc) was tuned to change the conductivity of YPtBi. The Co, Pt, MgAl$_2$O$_4$, and Ta layers were deposited on top of YPtBi at room temperature without breaking the vacuum.

**Device fabrication**

The samples were patterned into Hall bar structures with 180 μm-long × 60 μm-wide for DC planar Hall measurements, 90 μm-long × 25 μm-wide for second harmonic measurements, and 140 μm-long × 10 μm-wide for magnetization switching measurements by optical lithography and ion-milling techniques. A 50 nm-thick Ta and a 6 nm-thick Pt were deposited as electrodes by sputtering technique at room temperature, which reduces the effective length of the devices to 800 μm for DC planar Hall measurements, 50 μm for second harmonic measurements and 60 μm for magnetization switching measurements.

**Transport measurements**

The samples were mounted inside a vacuumed cryostat equipped with an electromagnet. For the DC planar Hall measurements, a Keithley 2400 Sourcemeter was used as the current source, and the Hall voltage was measured by using a Keithley 2002 Multimeter. For the second harmonic measurements, a NF LI5650 lock-in amplifier was employed to detect the first and the second harmonic Hall voltages under sine wave excitation generated by a Keithley 6221 AC/DC current source. For the DC (pulse) current-induced SOT magnetization switching, a Keithley 2400 SourceMeter (6221 AC/DC current source) was used, and the Hall signal was measured by a Keithley 2182A NanoVoltmeter. All experiments were performed at room temperature.

## Data availability

The data that support this study results are available from the corresponding author upon reasonable request.

## Acknowledgements

This work was supported by Kioxia corporation. The authors thank S. Miyajima Laboratory, S. Nakagawa Laboratory and Open Facility Center, Materials Analysis Division at Tokyo Institute of Technology for helps on XRD, SQUID and XRF measurements.

## Author contributions

T.S. fabricated the thin films and made the Hall devices. T.S. performed measurements with helps from T.F. and N.H.D.K.; T.K. and P.N.H. planned the experiments; T.S. and P.N.H. analyzed the data. T.S. and P.N.H. wrote the manuscript, with comments from T.K., T.F. and N.H.D.K.

## Competing interests

Tokyo Tech and Kioxia corporation have filed a joint patent basing on the finding in this work. All authors declare no further competing interests.

## References

[1] Cubukcu, M. *et al*. Spin-orbit torque magnetization switching of a three-terminal perpendicular magnetic tunnel junction. *Appl. Phys. Lett.* **104**, 042406 (2014).

[2] Fukami, S. *et al*. A spin–orbit torque switching scheme with collinear magnetic easy axis and current configuration. *Nat. Nanotech*. **11**, 621 (2016).

[3] Parkin, S. P. *et al*. Magnetic Domain-Wall Racetrack Memory. *Science* **320**, 190 (2008).

[4] Liu, L. *et al*. Magnetic Oscillations Driven by the Spin Hall Effect in 3-Terminal Magnetic Tunnel Junction Devices. *Phys. Rev. Lett.* **109**, 186602 (2012).

[5] Shirokura, T. and Hai, P. N. Bias-field-free spin Hall nano-oscillators with an out-of-plane precession mode. *J. Appl. Phys.* **127**, 103904 (2020).

[6] Liu, L. *et al*. Spin-torque switching with the giant spin Hall effect of tantalum. *Science* **336**, 555–558 (2012).

[7] Pai, C.-F. *et al*. Spin transfer torque devices utilizing the giant spin Hall effect of tungsten. *Appl. Phys. Lett.* **101**, 122404 (2012).

[8] Liu, L., Lee, O. J., Gudmundsen, T. J., Ralph, D. C. & Buhrman, R. A. Current-Induced Switching of Perpendicularly Magnetized Magnetic Layers Using Spin Torque from the Spin Hall Effect. *Phys. Rev. Lett.* **109**, 096602 (2012).

[9] Mellnik, A. R. *et al*. Spin-transfer torque generated by a topological insulator. *Nature* **511**, 449-451 (2014).

[10] Wu, H. *et al*. Room-Temperature Spin-Orbit Torque from Topological Surface States. *Phys. Rev. Lett.* **123**, 207205 (2019).

[11] Khang, N. H. D. *et al*. A conductive topological insulator with large spin Hall effect for ultralow power spin–orbit torque switching. *Nat. Mater.* **17**, 808-813 (2018).

[12] Chen, T., Peng, C., Tsai, T., Liao, W., Wu, C., Yen, H., and Pai, C. Efficient Spin−Orbit Torque Switching with Nonepitaxial Chalcogenide Heterostructures. *ACS Appl. Mater. Interfaces* **12**, 7788 (2020).

[13] Mahendra, DC. *et al*. Room-temperature high spin–orbit torque due to quantum confinement in sputtered $Bi_xSe_{1-x}$ films. *Nat. Mat.* **17**, 800–807 (2018).

[14] Khang, N. H. D., Nakano, S., Shirokura, T., Miyamoto, Y., Hai, P. N. Ultralow power spin–orbit torque magnetization switching induced by a non-epitaxial topological insulator on Si substrates. *Sci. Rep.* **10**, 12185 (2020).

[15] Fan, T., Khang, N. H. D., Shirokura, T., Huy, H. H., Hai, P. N. Low power spin–orbit torque switching in sputtered BiSb topological insulator/perpendicularly magnetized CoPt/MgO multilayers on oxidized Si substrate. *Appl. Phys. Lett.* **119**, 082403 (2021).

[16] Shi, S. *et al*. All-electric magnetization switching and Dzyaloshinskii-Moriya interaction in $WTe_2$/ferromagnet heterostructures. *Nat. Nanotechnol.* **14**, 945-949 (2019).

[17] Kimata, M., Chen, H., Kondou, K., Sugimoto, S., Muduli, P. K., Ikhlas, M., Omori, Y., Tomita, T., MacDonald, A. H., Nakatsuji, S. *et al.* Magnetic and magnetic inverse spin Hall effects in a non-collinear antiferromagnet. *Nature* **565**, 627–630 (2019).

[18] Chadov, S. *et al*. Tunable multifunctional topological insulators in ternary Heusler compounds. *Nat. Mater.* **9**, 541-545 (2010).

[19] Lin, H. *et al*. Half-Heusler ternary compounds as new multifunctional experimental platforms for topological quantum phenomena. *Nat. Mater.* **9**, 546-549 (2010).

[20] Al-Sawai, W. *et al*. Topological electronic structure in half-Heusler topological insulators. *Phys. Rev. B* **82**, 125208 (2010).

[21] Müchler, L. *et al*. Topological Insulators from a Chemist's Perspective. *Angew. Chem., Int. Ed. Engl.* **51**, 7221 (2012).

22  Nakajima, Y. *et al.* Topological *R*PdBi half-Heusler semimetals: A new family of noncentrosymmetric magnetic superconductors. *Sci. Adv*. **1**, e1500242 (2015).

23  Yan, B. and de Visser, A. Half-Heusler topological insulators. *MRS Bull.* **39**, 859-66 (2014).

24  Armitage, N. P. *et al*. Weyl and Dirac semimetals in three-dimensional solids. *Rev. Mod. Phys.* **90**, 015001 (2018).

25  Liu, Z. K. *et al.* Observation of unusual topological surface states in half-Heusler compounds LnPtBi (Ln = Lu, Y). *Nat. Commun.* **7**, 12924 (2016).

26  Logan, J. A. *et al.* Observation of a topologically non-trivial surface state in half-Heusler PtLuSb (001) thin films. *Nat. Commun.* **7**, 11993 (2016).

27  Kronenberg, A. *et al.* Dirac cone and pseudogapped density of states in the topological half-Heusler compound YPtBi. *Phys. Rev B* **94**, 161108(R) (2016).

28  Hosen, M. M. *et al.* Observation of Dirac state in half-Heusler material YPtBi. *Sci. Rep.* **10**, 12343 (2020).

29  Haase, M. G. *et al.* Equiatomic rare earth (Ln) transition metal antimonides LnTSb (T = Rh, Ir) and bismuthides LnTBi (T = Rh, Ni, Pd, Pt). *J. Solid State Chem.* **168**, 18–27 (2002).

30  Pavlosiuk, O. *et al.* Magnetic and Transport Properties of Possibly Topologically Nontrivial Half-Heusler Bismuthides RMBi (R = Y, Gd, Dy, Ho, Lu; M = Pd, Pt). *Acta Phys. Pol. A* **130**, 573 (2016).

31  Taskin, A. A. *et al.* Planar Hall effect from the surface of topological insulators. *Nat. Commun.* **8**, 1340 (2017).

32  Zhou, L. *et al.* Surface-induced positive planar Hall effect in topological Kondo insulator $SmB_6$ microribbons. *Phys. Rev. B* **99**, 155424 (2019).

[33] Bhardwaj, A. *et al.* Observation of planar Hall effect in topological insulator-$Bi_2Te_3$. *Appl. Phys. Lett.* **118**, 241901 (2021).

[34] Budhani, R. C., Higgins, J. S., McAlmont, D. and Paglione, J. Planar Hall effect in c-axis textured films of $Bi_{85}Sb_{15}$ topological insulator. *AIP Advances* **11**, 055020 (2021).

[35] Nguyen, M.-H. *et al.* Spin Torque Study of the Spin Hall Conductivity and Spin Diffusion Length in Platinum Thin Films with Varying Resistivity. *Phys. Rev. Lett.* **116**, 126601 (2016).

[36] Wang, Y. *et al.* Determination of intrinsic spin Hall angle in Pt. *Appl. Phys. Lett.* **105**, 152412 (2014).

[37] Khang, N. H. D. and Hai, P. N. Spin-orbit torque as a method for field-free detection of in-plane magnetization switching. *Appl. Phys. Lett.* **117**, 252402 (2020).

[38] Hayashi, M. *et al.* Quantitative characterization of the spin-orbit torque using harmonic Hall voltage measurements. *Phys Rev. B* **89**, 144425 (2014).

[39] Avci, C. O. *et al.* Interplay of spin-orbit torque and thermoelectric effects in ferromagnet/normal-metal bilayers. *Phys. Rev. B* **90**, 224427 (2014).

[40] Yang, H. *et al.* Characterization of spin-orbit torque and thermoelectric effects via coherent magnetization rotation. *Phys. Rev. B* **102**, 024427 (2020).

[41] Tanaka, T. *et al*. Intrinsic spin Hall effect and orbital Hall effect in 4*d* and 5*d* transition metals. *Phys. Rev. B* **77**, 165117 (2008).

[42] Sagasta, E. *et al.* Tuning the spin Hall effect of Pt from the moderately dirty to the superclean regime. *Phys. Rev. B* **94**, 060412(R) (2016).

[43] Zhang, Y. *et al.* Crossover of the three-dimensional topological insulator $Bi_2Se_3$ to the two-dimensional limit. *Nat. Phys.* **6**, 584-588 (2010).

[44] Liu, L. *et al.* Current-Induced Switching of Perpendicularly Magnetized Magnetic

Layers Using Spin Torque from the Spin Hall Effect. *Phys. Rev. Lett.* **109**, 096602 (2012).

[45] Jinnai, B., Zhang, C., Kurenkov, A., Bersweiler, M., Sato, H., Fukami, S., Ohno, H. Spin-orbit torque induced magnetization switching in Co/Pt multilayers. *Appl. Phys. Lett.* **111**, 102402 (2017).

[46] Coffey, W. T. and Kalmykov, Y. P. Thermal fluctuations of magnetic nanoparticles: Fifty years after Brown. *J. Appl. Phys.* **112**, 121301 (2012).

[47] Koch, R. H. *et al.* Time-resolved reversal of spin-transfer switching in a nanomagnet. *Phys. Rev. Lett.* **92**, 088302 (2004).

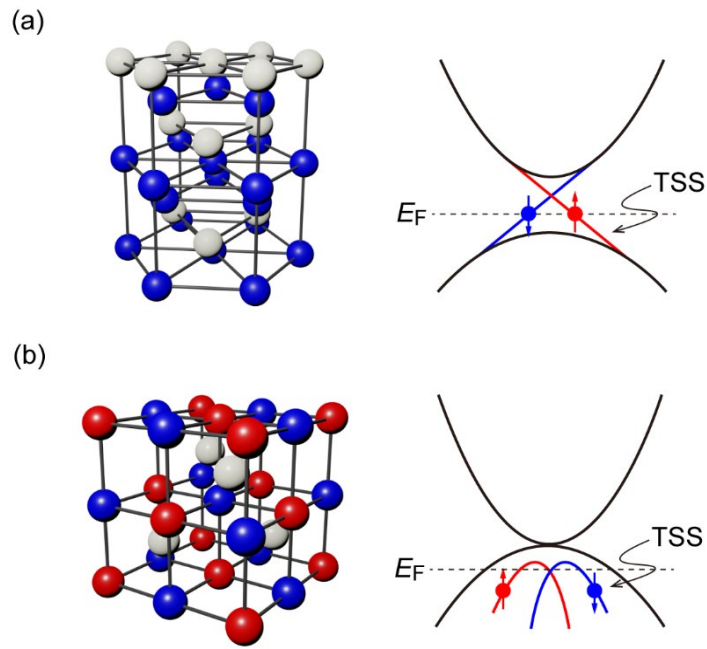

**Figure 1.** Comparison of the crystal structure and band structure of (a) topological insulator and (b) half-Heusler alloy topological semimetal. In the band structures, TSS indicates topological surface states.

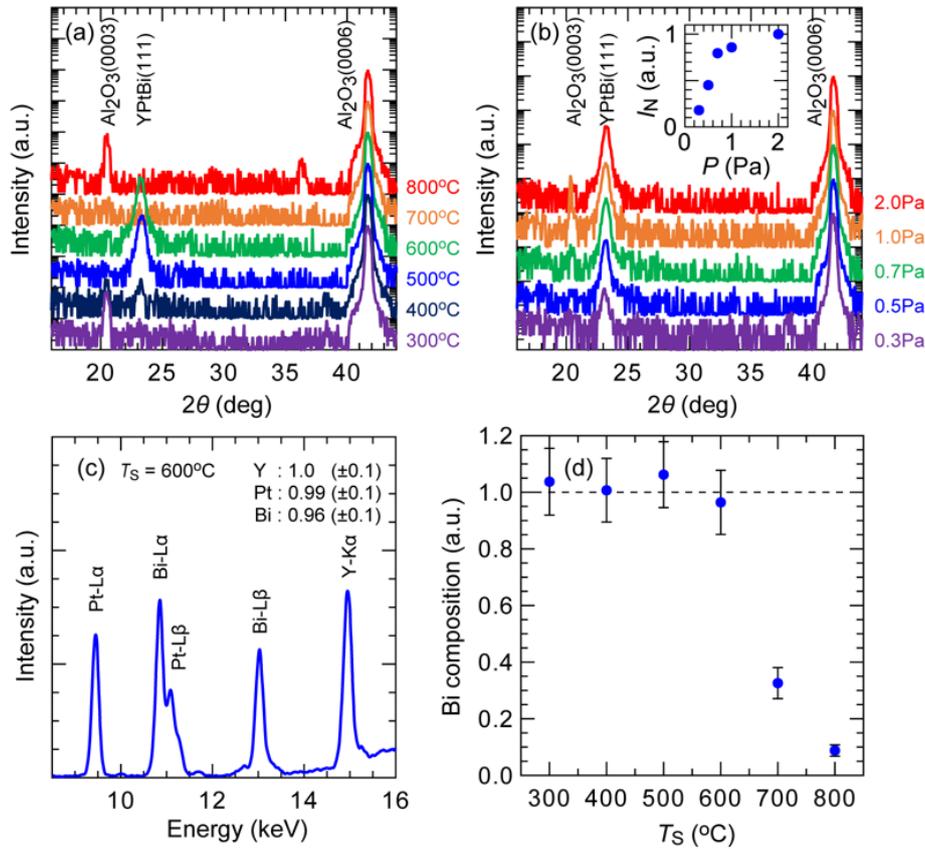

**Figure 2. Crystal structure analysis of YPtBi thin films. (a)** X-ray diffraction (XRD) $\theta$-$2\theta$ spectra of 50 nm-thick YPtBi films deposited on c-sapphire at different substrate temperature ranging from 300°C to 800°C and the Ar pressure of 0.2 Pa. **(b)** XRD $\theta$-$2\theta$ spectra of YPtBi films deposited at different Ar pressure ranging from 0.3 to 2.0 Pa at the substrate temperature of 600°C. Inset shows the peak intensity of YPtBi(111) normalized by that at 2.0 Pa as a function of Ar pressure. **(c)** X-ray fluorescence spectroscopy (XRF) of an YPtBi thin film deposited at 600°C. **(d)** Bi composition at different substrate temperature. YPtBi is stable up to 600°C.

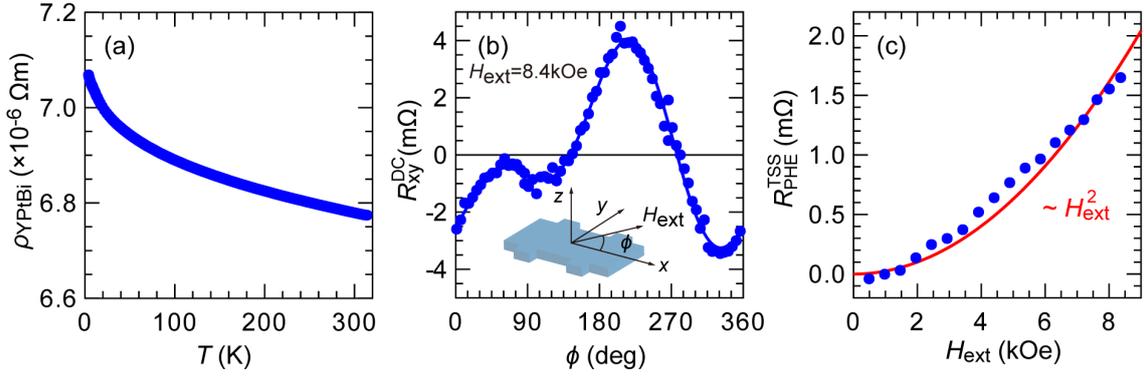

**Figure 3. (a) Temperature dependence of the resistivity of a 11.8 nm-thick YPtBi film. (c) Planar Hall resistance measured with an in-plane rotating $H_{ext}$ of 8.4 kOe measured at 4K, where dots are experimental data and the solid curve is a fitting given by equation (1). Inset shows the coordination system and definition for the azimuth angle $\phi$ of $H_{ext}$. The sin2$\phi$ dependence of the Hall resistance originates from the spin-momentum locking of TSS. (d) Amplitude of the sin2$\phi$ planar Hall resistance as a function of $H_{ext}$, which increases as $H_{ext}^2$, similar to those observed in TIs.**

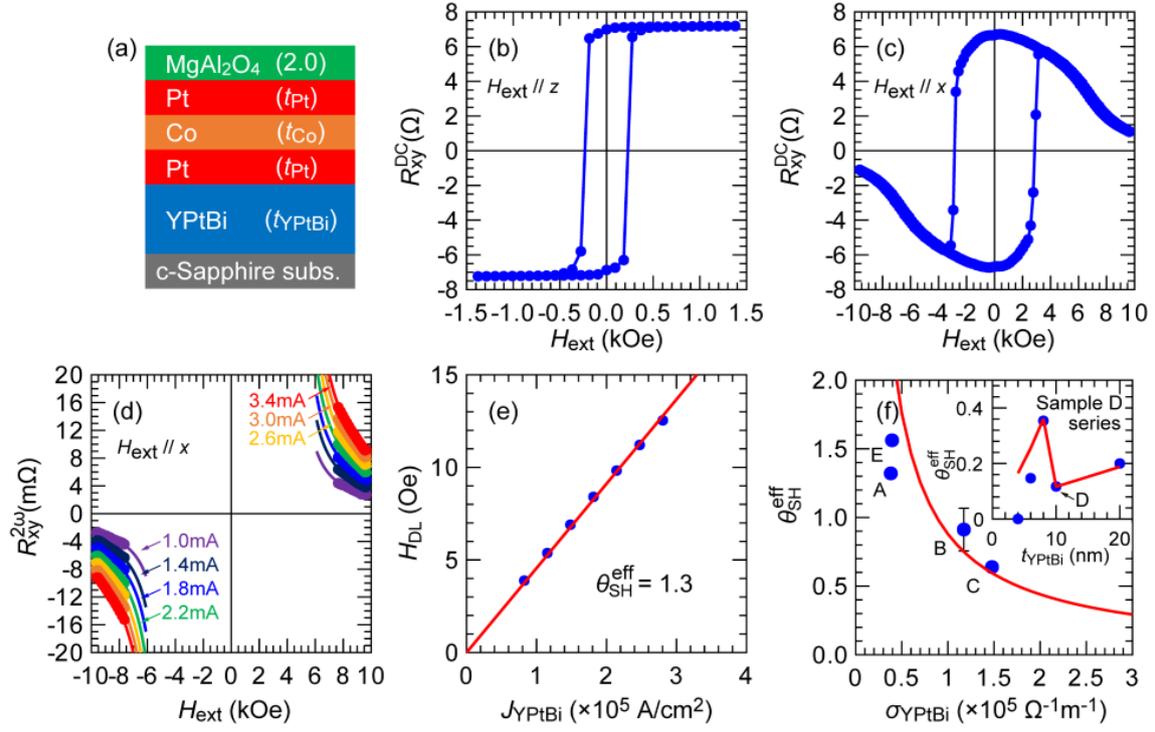

**Figure 4.** (a) Schematic illustration of sample structure for spin Hall effect measurements, where layer thicknesses are in nm. (b),(c) DC anomalous Hall resistance for sample A ($\sigma_{YPtBi} = 0.38\times10^5\ \Omega^{-1}m^{-1}$ and $t_{Co}=0.5$ nm) measured with $H_{ext}$ applied along the $z$-direction and the $x$-direction, respectively. (d) Second harmonic Hall resistance of sample A measured with $H_{ext}$ applied along the $x$-direction and an AC current ranging from 1.0 to 3.4 mA, where dots are experimental data and solid curves show fitting results given by equation (2). (e) Antidamping-like field $H_{DL}$ as a function of the current density in the YPtBi layer. (f) Relationship between the effective spin Hall angle and the conductivity of the YPtBi layer observed in sample A, sample B ($\sigma_{YPtBi} = 1.2\times10^5\ \Omega^{-1}m^{-1}$ and $t_{Co}=0.5$ nm), sample C ($\sigma_{YPtBi} = 1.5\times10^5\ \Omega^{-}$

$^{1}$m$^{-1}$ and $t_{Co}$=0.8 nm), and sample E ($\sigma_{YPtBi}$ = 0.39×10$^5$ Ω$^{-1}$m$^{-1}$ and $t_{Co}$=0.5 nm). Inset shows the thickness dependence of the effective spin Hall angle in another series of samples. Sample D ($\sigma_{YPtBi}$ = 2.0×10$^5$ Ω$^{-1}$m$^{-1}$ and $t_{Co}$=0.5 nm) is a representative sample in this series.

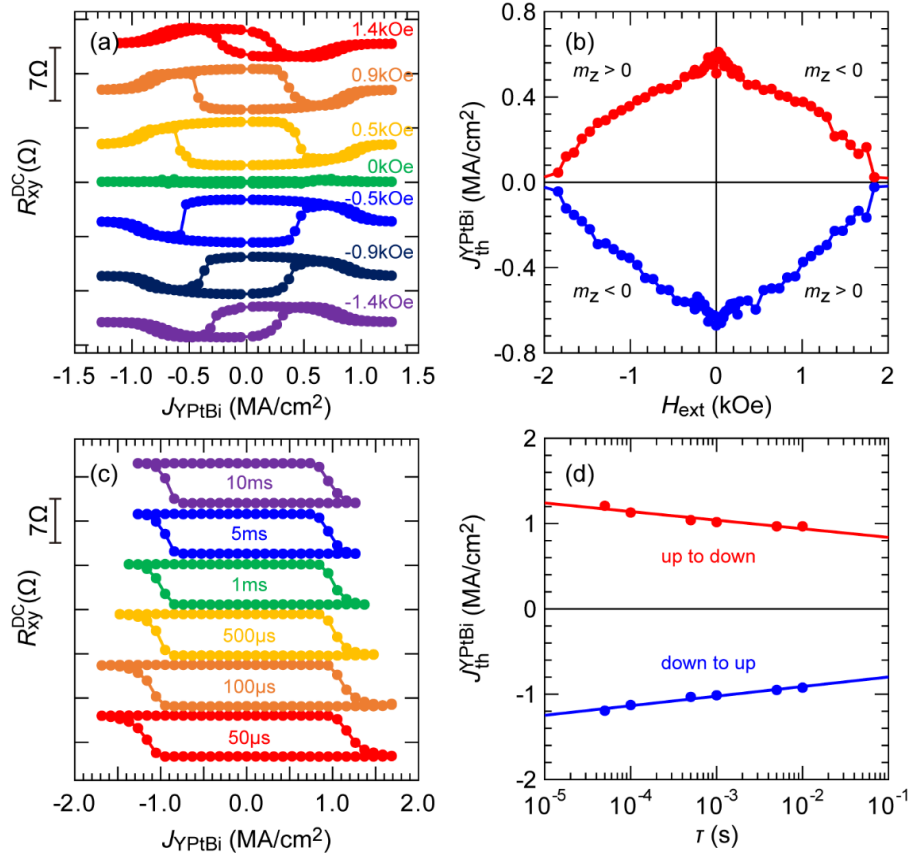

**Figure 5.** Ultralow current-induced magnetization switching in sample E ($\sigma_{YPtBi}$ = 0.39×10$^5$ Ω$^{-1}$m$^{-1}$ and $t_{Co}$=0.5 nm). (a) SOT magnetization switching by DC currents with various in-plane $H_{ext}$ (-1.4 to 1.4 kOe) applied along the *x*-direction. (b) Threshold switching current density in YPtBi as a function of $H_{ext}$. (c) SOT magnetization switching by pulse currents with various pulse width ranging from 50 μs to 10 ms and $H_{ext}$ of 0.5 kOe. (d) Threshold switching current density in YPtBi as a function of pulse width, where dots are experimental data and solid lines show fitting results given by equation (5).

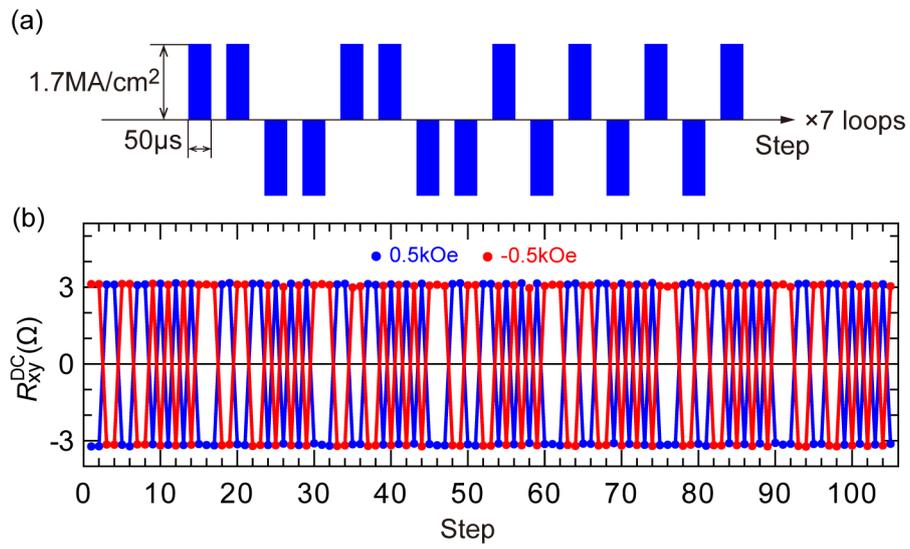

**Figure 6. Robust SOT magnetization switching by YPtBi. (a) Sequence of 105 pulses with the current density in YPtBi of $1.7\times10^6$ Acm$^{-2}$ and the pulse width of 50 μs. (b) Hall resistance data measured for a total of 210 pulses under a bias field of ±0.5 kOe.**